# Large electrobending deformation caused by defect dipoles

Shuo Tian, Bin Li*, Yejing Dai*

School of Materials, Sun Yat-sen University, Shenzhen, 518107, P. R. China
*Corresponding authors.
Email: libin75@mail.sysu.edu.cn (B.L.); daiyj8@mail.sysu.edu.cn (Y.D.)

**Abstract**

Ultrahigh electrostrains (> 1%) in several piezoceramic systems have been reported since 2022, which attract more and more interest in the field of piezoelectricity; however, the mechanism is still unclear. Here, we have directly observed a novel electric field-induced bending (electrobending) phenomenon that visually exhibits as an alternating concave-convex deformation under an electric field of ±50 kV cm$^{-1}$, in nonstoichiometric $(K_{0.48}Na_{0.52})_{0.99}NbO_{2.995}$ ceramics, which causes the measured ultrahigh electrostrain. It is demonstrated that the electrobending deformation arises from the different stresses due to the stretching or compression of the defect dipoles on the upper and lower surfaces of the ceramics. As a result of the large electrobending deformation, a giant apparent electrostrain of 11.6% is obtained at room temperature, and it can even reach up to 26.0% at 210 °C, which far exceeds that of all present piezoelectric materials. Our discovery is an important addition and refinement to the field of condensed matter physics, whilst providing a new strategy and shedding light on the design of future precision actuators or intelligent devices.

**Introduction**

Piezoelectric materials, which transform electrical energy into mechanical energy, are used as actuators in a variety of technological applications[1-4]. The use of actuators requires materials that experience strain in response to an applied electric field[5]. Over the past few decades, significant research into perovskite ferroelectrics has produced a number of materials with outstanding electrostrain properties. Composition design[6, 7] (e.g., morphotropic phase boundary), engineered domain configuration[8, 9], electric-field-induced phase transition[10, 11], chemical disorder[12], reversible switching of the non-180º domains[13, 14], and nanocomposite structures[15, 16] have all been explored to achieve a large electric-field-induced strain (electrostrain) response in piezoelectrical materials. All of these approaches show that it is possible to increase the piezoelectric response by an order of magnitude (~1%) in comparison to that of the commercial Pb(Zr, Ti)O$_3$ (PZT) ceramics (~ 0.1%)[6]. These electrostrains originate from small deformations of the crystal unit cell (lattice strain) under an applied electric field. The associated displacements of ions are in the picometer range, and thus few electrostrain values of more than 1% in piezoelectrical materials are reported due to the limitation of the ion displacement distance.

Since 2022, reports on the ultrahigh electrostrains (> 1%) in several piezoceramics have been spurred, constantly breaking records, and defect dipoles are thought to be the main contributor to the ultrahigh strain in these systems[17-23]. Interestingly, most of these ultrahigh electrostrains are achieved when the thickness of the samples is below 300 μm. Recently, Ranjan *et al.* reported that when the thickness of piezoceramic disks (10-12 mm in diameter) is less than 500 μm, the measured electrostrain values shoot up sharply[24]. The electrostrain of Na$_{0.5}$Bi$_{0.5}$TiO$_3$ and K$_{0.5}$Na$_{0.5}$NbO$_3$-based lead-free systems greatly increased with decreasing thickness and could obtain much higher strain levels (4% - 5%) in the small thickness regime (< 300 μm), which is difficult to achieve only through lattice strain. Though the exact cause of this ultrahigh electrostrain is yet unknown, the electric field-induced mechanical deformation is certainly related to the presence of defect dipoles.

In this study, a huge electrostrain of > 10% at room temperature and > 25% at 210 °C is achieved in $(K_{0.48}Na_{0.52})_{0.99}NbO_{2.995}$ piezoceramic samples with a thickness of 140 μm and a diameter of 10 mm. In order to investigate the origin of the large electric-field-induced deformation of the samples, we used a laser scanning vibrometer with an AC bipolar electric field of 50 kV cm$^{-1}$ in magnitude at 1 Hz for the KNN ceramic sample. By obtaining the displacement change at each point, the real-time vibration process of the samples is successfully observed. When the sample thickness is relatively large (> 500 μm), the sample behaves as a rigid material with a small deformation, which is mainly from the lattice strain, showing the typical traditional piezoelectric characteristics. When the thickness of the sample is reduced to a certain extent (< 300 μm), the sample shows an electrobending (electric-field-induced bending) characteristic, which contributes to the large apparent electrostrain. This phenomenon well explains the effect of the thickness on the electrostrain, since the thinner the sample, the easier it is to bend. Further investigations reveal a reversible bending deformation mechanism induced by oriented-defect dipoles in the upper and lower surface layers of the ceramic samples. This new discovery is an important addition and refinement to the field of condensed matter physics, while also providing a new strategy and shedding light on the design of future actuators.

**Results and discussion**

Due to the crucial effect of defect dipoles on performance, we chose the non-stoichiometric $(K_{0.48}Na_{0.52})_{0.99}NbO_{2.995}$ (KNN) composition in order to form <110>-oriented $(V'_A - V^{\cdot\cdot}_O)$ defect dipoles more effectively. X-ray diffraction (XRD) was carried out to demonstrate the formation of an orthorhombic phase with the perovskite structure of KNN ceramics (Fig. S1). The strain-electric field (*S-E*) curves of this composition for different sample thicknesses and electric fields are shown in Figs. 1A, 1B, and S2, and the polarization-electric field (*P-E)* loops are plotted in Fig. S3. Obviously, the electrostrain value increases as the sample thickness decreases; especially when the sample thickness is less than 300 μm, the bipolar electrostrain values increase dramatically, escalating from 0.06% with a 1.0 mm thickness to 11.6% with a 140 μm thickness at an electric field of -50 kV cm$^{-1}$. Moreover, as the sample thickness decreases, the bipolar *S-E* curve gradually becomes asymmetric and is no longer a typical butterfly shape (Fig. 1A), which is similar to the results of previously reported work[24].

Furthermore, the electrostrain of KNN ceramics is not only related to the thickness but also to the dimension of the sample holder. As the dimension of the bottom sample holder increases, the value of the measured electrostrain gradually increases from 0.2% with a 1.0 mm diameter to 11.6% with a 10 mm diameter (Fig. S4). Therefore, we infer that the sample is undergoing a nonhomogeneous deformation under a uniform electric field, which induces the asymmetrical bipolar *S-E* curves (Fig. 1A).

For comparison, the electrostrain value of BaTiO$_3$ (BT) piezoelectric ceramics has little dependence on sample thickness (Fig. S5), only increasing from 0.17% with a 1 mm thickness to 0.18% with a 200 μm thickness. The reason for this divergence is that there are few defect dipoles created in the BT sample due to the non-volatile nature of cations (Ba$^{2+}$ and Ti$^{4+}$) when sintered in the air. These results further confirm that the giant electric-field-induced deformation in KNN ceramics originates from defect dipoles.

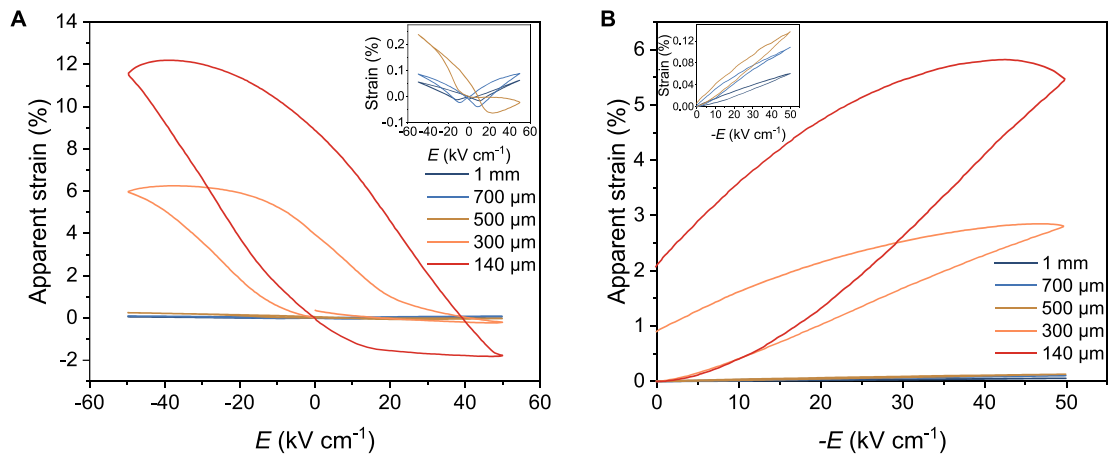

**Fig. 1. Strain performance of KNN ceramics.** (**A**) Bipolar and (**B**) unipolar *S-E* curves of KNN ceramics with different thicknesses.

To obtain the real-time deformation of KNN ceramics under an electric field, we monitored the vibration of KNN ceramics with a thickness of 140 μm and a diameter of 10 mm under an AC bipolar field of 50 kV cm$^{-1}$ in magnitude at 1 Hz using a laser scanning vibrometer. Figure 2A illustrates the displacement of KNN ceramics under an electric field of ±50 kV cm$^{-1}$, and an obvious bending can be observed. The edge of the ceramic exhibits upward bending under a positive electric field, and the maximum displacement at the edge reaches 18 μm, which is equivalent to an apparent strain of 12.9% along the thickness direction compared to the zero electric field state. When the electric field is reversed, the sample bends downward by almost the same amount accordingly. It is worth noting that the ceramics always bend towards the surface connected to the positive electrode, as shown in Fig. S6.

Theoretically speaking, such a deformation should lead to a symmetric bipolar *S-E* curve. However, the obtained bipolar *S-E* curves are asymmetric (Fig. 1A), and giant electrostrains can only be observed under the negative electric field. This can be attributed to the limitations of currently commonly used measurement equipment (aix-ACCTF Analyzer 2000E). Generally, the bottom sample holder is a platform with a relatively larger diameter (3 mm), and the top sample holder is a probe with a curved tip (1 mm), as shown in Fig. S7. The displacement distance of the top sample holder determines the obtained displacement signal. Therefore, only when the sample bends downward can the top sample holder be effectively lifted, leading to a giant apparent strain. When the sample bends upward, the top sample holder cannot be effectively lifted, leading to a small or even negative strain (Fig. S7). These phenomena also explain why the strain value is related to the diameter of the bottom sample holder (Fig. S4). Therefore, the direction of asymmetry of the *S-E* curve can be changed by changing the dimensions of the top and bottom sample holders (Fig. S8); however, this cannot be realized by turning the sample over (Fig. S9).

The deformation of 500 μm-thick KNN ceramics is also investigated (Fig. 2B), and there is no obvious bending observed. The maximum displacement is only 1.9 μm, which corresponds to a strain of 0.3%. This is due to the fact that the sample with a larger thickness does not bend easily, whereas the sample with a relatively smaller thickness is easier to bend, which is exactly why the strain value in ceramics is sensitive to the thickness of the samples. As a comparison, the deformation of the pure BT ceramic sample with a thickness of 200 μm is plotted in Fig. 2C, and a maximum displacement of 0.32 μm at the edge can be obtained, which is further evidence that the generation of electrobending is closely related to defect dipoles.

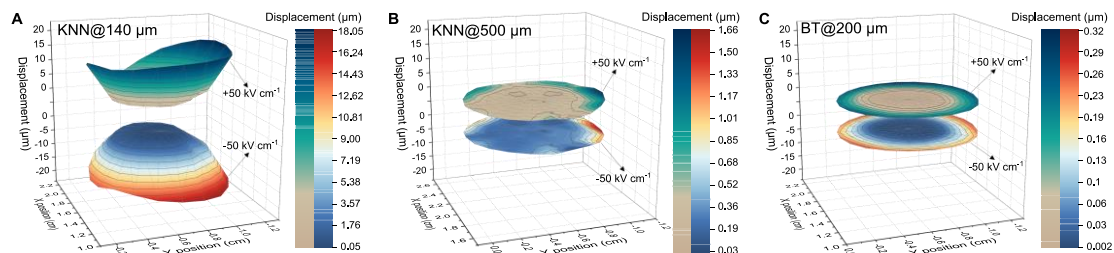

**Fig. 2. Electrobending deformation of KNN ceramics.** 3D drawing of surface displacement of KNN ceramics with a thickness of (**A**) 140 μm and (**B**) 500 μm. (**C**) 3D drawing of surface displacement of BT ceramics with a thickness of 200 μm.

Why do piezoelectric ceramics with small thicknesses undergo this special bending deformation under an electric field? A working mechanism is proposed to explain the electrobending effect in piezoelectric materials, as shown in Figs. 3A-3C. The A-site vacancies ($V'_K$ and $V'_{Na}$) and oxygen vacancies ($V^{\cdot\cdot}_O$) are introduced into KNN ceramics by the nonstoichiometric composition design. We suggest that some unanticipated <110>-oriented ($V'_A - V^{\cdot\cdot}_O$) defect dipoles could be formed in some unit cells during the sintering and subsequent preparation processes. After aging for some time in the ferroelectric state, the number of defect dipoles further increases *via* the diffusion of oxygen vacancies. During the preparation process, defects are more likely to occur on the surface of the ceramics compared with the inside of the ceramic. The results of X-ray photoelectron spectroscopy (XPS) demonstrate the difference in oxygen vacancy content between the surface and inside of the ceramic (Fig. 3D). The shoulder peaks of the O1s orbitals are generally believed to be generated by adsorbed oxygen due to oxygen vacancies[25, 26]. The intensity of the shoulder peaks in the ceramic surface is higher than that in the ceramic inside, indicating a higher concentration of oxygen vacancies in the surface layer. Therefore, it is easier to form defect dipoles in the surface layer compared with the inside of ceramics[17].

The surface layer of piezoelectric ceramics is in a more unstable and high-energy environment than the inside. In order to reduce the surface energy, the defect dipoles of the surface layer are more inclined to be oriented inward to reduce the energy through internal electric dipole interactions (Fig. 3A). Consequently, the orientation of most spontaneous polarizations near defect dipoles in the surface layer in KNN ceramics also faces inward, based on "the symmetry-conforming property of point defects"[27]. There is a strong interaction between defect dipoles and spontaneous polarizations. Figure 3E shows the results of local poling experiments with piezoresponse force microscopy (PFM) for KNN ceramic surface. When applying a ±150 V external voltage to the different concentric square-shaped regions, respectively, there is no obvious domain switching behavior observed due to the pinning effect of the defect dipoles to the ferroelectric domain[28, 29], further confirming the existence of defect dipoles in the surface layer.

The reorientation of the defect dipole involves an energy-consuming diffusion process, which requires the migration of vacancies, and thus it becomes challenging to switch the defect dipoles at a relatively low applied electric field. Therefore, the defect dipoles will be stretched or compressed when subjected to an electric field in different directions[27, 30]. As shown in Fig. 3B, the defect dipoles and spontaneous polarizations in the upper surface will be compressed along the thickness direction when an upward electric field is applied, leading to an expansion in the in-plane direction. Meanwhile, the defect dipoles and spontaneous polarizations in the lower surface will be stretched along the thickness direction, leading to a contraction in the in-plane direction. Consequently, the asymmetric strain states of the upper and lower surfaces in KNN ceramics under an electric field cause downward bending deformation. When the electric field is reversed, the ceramics will bend upward (Fig. 3C). Since the response of the defect dipole under the electric field is dynamic and reversible, when the electric field is withdrawn, the defect dipole will return to its initial state, causing the ceramic to revert to a flat state.

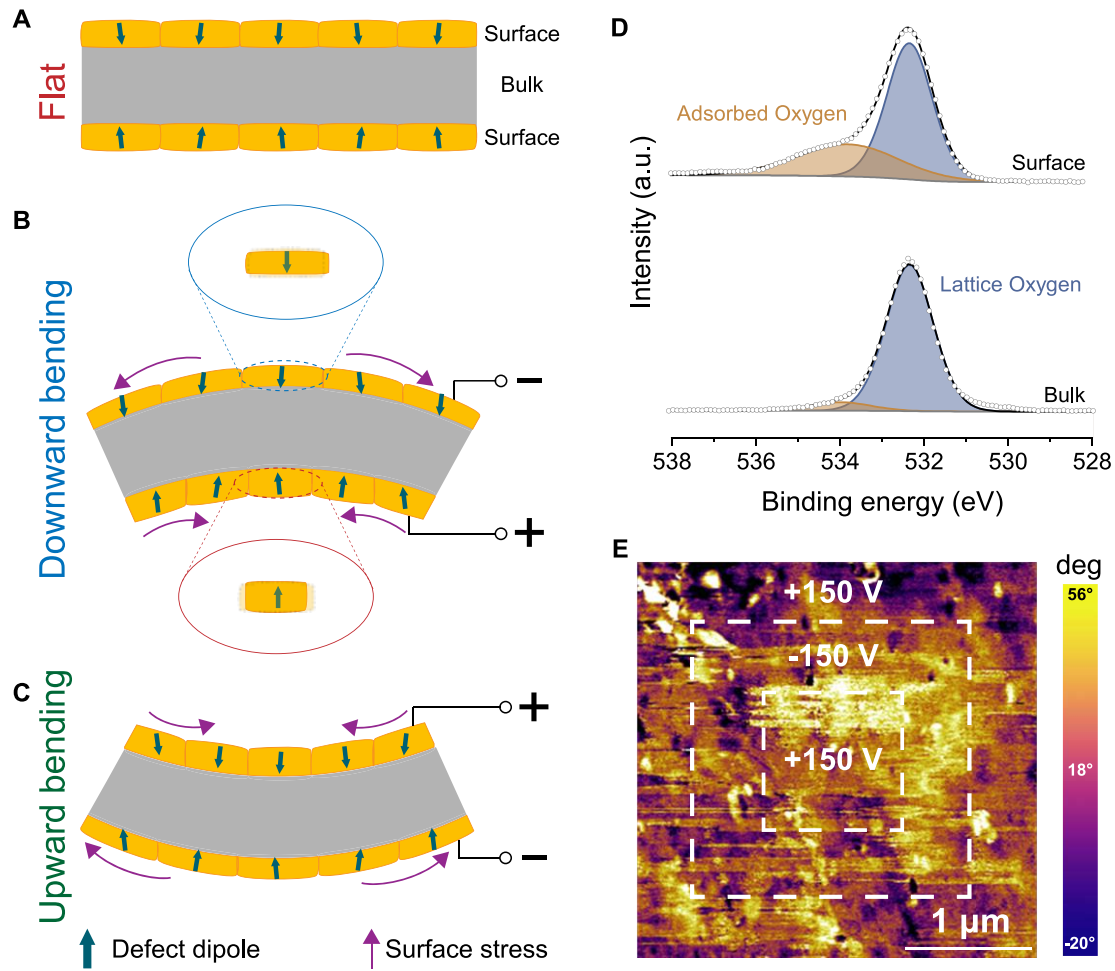

**Fig. 3. Working mechanism for electrobending.** (**A-C**) Schematic diagram of electrobending in KNN ceramics with defect dipoles under an electric field. (**D**) X-ray photoelectron spectroscopy (XPS) of O1s orbitals. (**E**) PFM characterization of KNN ceramic surface after writing domain measurement.

In order to better evaluate the electrobending performance of KNN ceramics as actuators, the cycling reliability and temperature stability tests are investigated, as shown in Figs. 4, S10 and S11. After $10^6$ cycles, the strain value decreases by only 4.3%, as shown in Figs. 4A and 4B, suggesting good fatigue-resistant behavior. This is because the large electrobending strain is little related to the domain switching or phase transition mechanism[31-33]. Furthermore, the bending strain increases significantly with increasing temperature (Figs. 4B and S11) and reaches its maximum value of 26.0% at 210 °C, due to the fact that the temperature is at the orthorhombic-tetragonal phase transition (Fig. S12), and higher temperatures make ceramics with less stiffness and thus more flexible to bend. The above excellent performance suggests that the electric field-induced bending will open a new field for piezoelectric actuators.

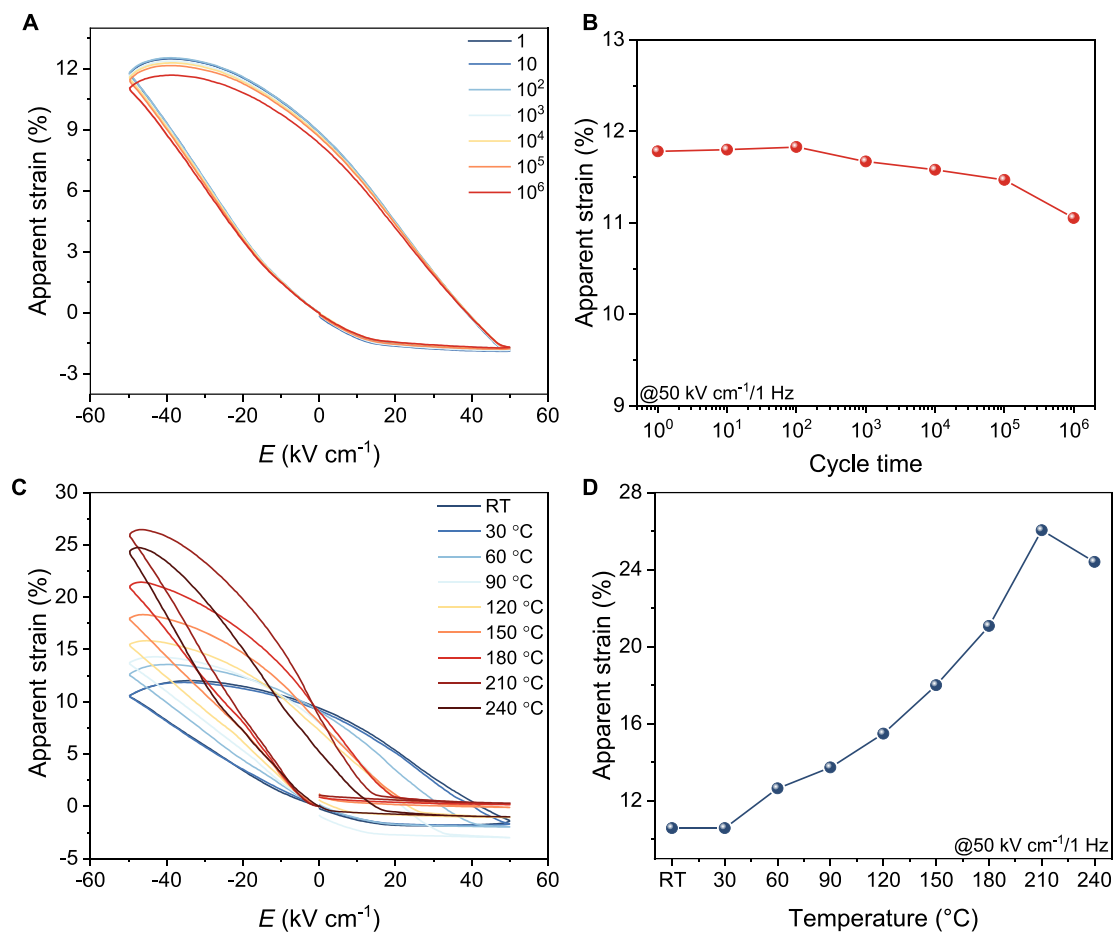

**Fig. 4. Fatigue behavior and temperature stability of electrobending.** (**A**) and (**B**) Fatigue tests of *S-E* curves under an electric field of 50 kV cm$^{-1}$. (**C**) and (**D**) Change of bipolar *S-E* curves from 30 °C to 240 °C.

**Conclusions**

In summary, we report a novel electrobending effect, which reveals the nature of the phenomenon of the recently reported ultrahigh electrostrain values. The electrobending deformation is directly observed by the laser scanning vibrometry technology, and it exhibits an alternating concave-convex deformation with the change of electric field, arising from the different stresses due to the stretching or compression of the defect dipoles on the upper and lower surfaces of the ceramic. Due to the large electrobending deformation, a giant apparent electrostrain of 11.6% at room temperature and 26.0% at 210 °C is obtained. Giant displacements, good fatigue stability, and excellent performance at high temperatures indicate that the electrobending effect has great potential for the applications of intelligent devices or regulators in a variety of environments.

**Funding:**

National Key R&D Program of China (Grant No. 2021YFA0716500)

National Natural Science Foundation of China (52172135)

Youth Top Talent Project of the National Special Support Program (2021-527-07)

Leading Talent Project of the National Special Support Program (2022WRLJ003)

Guangdong Basic and Applied Basic Research Foundation for Distinguished Young Scholars (Grant No. 2022B1515020070 and 2021B1515020083)

**Data and materials availability:**
All data are available in the main text or the supplementary materials.

**Supplementary Materials**

Materials and Methods

Figs. S1 to S12

**References**

1. Tan, Z., et al., *New High-Performance Piezoelectric: Ferroelectric Carbon-Boron Clathrate.* Phys. Rev. Lett., 2023. **130**(24): p. 246802.
2. Zhang, S., et al., *Advantages and challenges of relaxor-PbTiO$_3$ ferroelectric crystals for electroacoustic transducers–A review.* Prog. Mater. Sci., 2015. **68**: p. 1-66.
3. Scott, J., *Applications of modern ferroelectrics.* Science, 2007. **315**(5814): p. 954-959.
4. Curry, E.J., et al., *Biodegradable nanofiber-based piezoelectric transducer.* Proceedings of the National Academy of Sciences, 2020. **117**(1): p. 214-220.
5. Hao, J., et al., *Progress in high-strain perovskite piezoelectric ceramics.* Mat. Sci. Eng. R., 2019. **135**: p. 1-57.
6. Saito, Y., et al., *Lead-free piezoceramics.* Nature, 2004. **432**(7013): p. 84-87.
7. Liu, H., et al., *Role of reversible phase transformation for strong piezoelectric performance at the morphotropic phase boundary.* Phys. Rev. Lett., 2018. **120**(5): p. 055501.
8. Finkel, P., et al., *Simultaneous Large Optical and Piezoelectric Effects Induced by Domain Reconfiguration Related to Ferroelectric Phase Transitions.* Adv. Mater., 2022. **34**(7): p. 2106827.
9. Narayan, B., et al., *Electrostrain in excess of 1% in polycrystalline piezoelectrics.* Nat. Mater., 2018. **17**(5): p. 427-431.
10. Wu, J., et al., *Ultrahigh field-induced strain in lead-free ceramics.* Nano Energy, 2020. **76**: p. 105037.

11. Zhang, Y., et al., *Simultaneous Realization of Good Piezoelectric and Strain Temperature Stability via the Synergic Contribution from Multilayer Design and Rare Earth Doping.* Adv. Funct. Mater., 2023. **33**(11): p. 2211439.

12. Cheng, M., et al., *Optimizing piezoelectric performance of complex perovskite through increasing diversity of B-site cations.* Journal of Materials Science & Technology, 2024. **170**: p. 78-86.

13. Habib, M., et al., *Design and development of a new lead-free $BiFeO_3$-$BaTiO_3$ quenched ceramics for high piezoelectric strain performance.* Chem Eng J, 2023. **473**: p. 145387.

14. Sun, X.-x., et al., *Understanding the piezoelectricity of high-performance potassium sodium niobate ceramics from diffused multi-phase coexistence and domain feature.* J Mater Chem A, 2019. **7**(28): p. 16803-16811.

15. Li, F., et al., *Giant piezoelectricity of Sm-doped Pb $(Mg_{1/3}Nb_{2/3})$ $O_3$-$PbTiO_3$ single crystals.* Science, 2019. **364**(6437): p. 264-268.

16. Li, J., et al., *Lead zirconate titanate ceramics with aligned crystallite grains.* Science, 2023. **380**(6640): p. 87-93.

17. Feng, W., et al., *Heterostrain-enabled ultrahigh electrostrain in lead-free piezoelectric.* Nat. Commun., 2022. **13**(1): p. 5086.

18. Huangfu, G., et al., *Giant electric field–induced strain in lead-free piezoceramics.* Science, 2022. **378**(6624): p. 1125-1130.

19. Luo, H., et al., *Achieving giant electrostrain of above 1% in (Bi, Na) $TiO_3$-based lead-free piezoelectrics via introducing oxygen-defect composition.* Sci. Adv., 2023. **9**(5): p. eade7078.

20. Lai, L., et al., *Giant Electrostrain in Lead-free Textured Piezoceramics by Defect Dipole Design.* Adv. Mater., 2023: p. 2300519.

21. Wang, B., et al., *Giant Electric Field-Induced Strain with High Temperature-Stability in Textured KNN-Based Piezoceramics for Actuator Applications.* Adv. Funct. Mater., 2023: p. 2214643.

22. Jia, Y., et al., *Giant electro-induced strain in lead-free relaxor ferroelectrics via defect engineering.* J. Eur. Ceram. Soc., 2023. **43**(3): p. 947-956.

23. Li, W., et al., *Giant electro-strain nearly 1% in $BiFeO_3$-based lead-free piezoelectric ceramics through coupling morphotropic phase boundary with defect engineering.* Mater. Today. Chem., 2022. **26**: p. 101237.

24. Adhikary, G.D., et al., *Ultrahigh electrostrain> 1% in lead-free piezoceramics: Role of disk dimension.* J. Appl. Phys., 2023. **134**(5).

25. Feng, Y., et al., *Defects and aliovalent doping engineering in electroceramics.* Chem. Rev., 2020. **120**(3): p. 1710-1787.

26. Lai, L., et al., *Ultrahigh electrostrain with excellent fatigue resistance in textured $Nb^{5+}$-doped $(Bi_{0.5} Na_{0.5}) TiO_3$-based piezoceramics.* Journal of Advanced Ceramics, 2023. **12**(3): p. 487-497.

27. Ren, X., *Large electric-field-induced strain in ferroelectric crystals by point-defect-mediated reversible domain switching.* Nat. Mater., 2004. **3**(2): p. 91-94.

28. Rodriguez, B.J., et al., *Ferroelectric domain wall pinning at a bicrystal grain boundary in bismuth ferrite.* Appl. Phys. Lett., 2008. **93**(14).

29. Zhao, Z.-H., Y. Dai, and F. Huang, *The formation and effect of defect dipoles in lead-free piezoelectric ceramics: a review.* Sustain. Mater. Techno., 2019. **20**: p. e00092.

30. Zhao, Z., et al., *Ultrahigh electro-strain in acceptor-doped KNN lead-free piezoelectric ceramics via defect engineering.* Acta Mater., 2020. **200**: p. 35-41.
31. Wu, B., et al., *Facilitated reversible domain switching at multiphase boundary in periodically orthogonal poled KNN-based ceramics: Strain versus non-180º domain.* J. Am. Ceram. Soc., 2023.
32. Sun, S., et al., *Role of tetragonal distortion on domain switching and lattice strain of piezoelectrics by in-situ synchrotron diffraction.* Scripta Materialia, 2021. **194**: p. 113627.
33. Kumar, N., et al., *Factors influencing the coupling between non-180° domain switching and lattice strain in perovskite piezoceramics.* Phys. Rev. B, 2018. **97**(13): p. 134113.

## Supplementary Materials

**Methods**

Sample preparation

The $(K_{0.48}Na_{0.52})_{0.99}NbO_{2.995}$ and $BaTiO_3$ ceramics were fabricated by a conventional high-temperature solid-state sintering method. $Na_2CO_3$ (99.8%), $K_2CO_3$ (99.5%), $Nb_2O_5$ (99.5%, orthorhombic phase), $BaCO_3$ (99.8%), and $TiO_2$ (99.5%) powders were selected as raw materials. The raw materials were mixed and ball-milled with yttrium stabilized zirconia balls in ethanol for 12 h. The mixed powders were calcined at 850 °C for 3 h in the air, followed by a ball mill for another 12 h. Then the calcined powders were uniaxially pressed into pellets with a diameter of 10 mm and a thickness of 1 mm under a pressure of 200 MPa. Polyvinyl butyral (PVB) was added as a binder to assist mold and burn off at 500 °C after pressing. The samples were sintered in air at 1100 °C for 4 h and buried in the self-source powders. Samples are polished and thinned to 140 μm ~ 1 mm and fired with silver on both sides as electrodes for electrical characteristic measurement.

Electrical measurements

The strain-electric field ($S$-$E$) curves were measured by a ferroelectric tester (aix-ACCTF Analyzer 2000E, Aachen Germany) equipped with a high-voltage amplifier (Trek 610E, TREK, USA) at a frequency of 1 Hz. A temperature controller was used to produce in-situ thermal field measurements in electrostrain tests. The fatigue resistance was measured at 100 Hz and tested at 1 Hz. The out-of-plane displacement was determined by a laser scanning vibrometer (PolyTech MSA-600, GmbH, Germany), which was excited with AC voltage generated by a signal generator connected through a power amplifier and monitored by an oscilloscope.

Structure characterizations

The phase structure of the samples was determined by X-ray diffraction analysis (D8 ADVANCE, Bruker, Germany) with Cu K$\alpha$ radiation. The sample surface was etched to 3 nm by Ar ions and then tested by X-ray photoelectron spectroscopy (Escalab 250Xi,

Thermo Fisher, Britain) equipped with a standard monochromatic AlK*a* excitation source ($hv$ = 1361 eV). The local poling experiment was carried out on an atomic force microscope (AFM) with the functionality of a piezo-force microscope (PFM, Cypher S, Asylum Research, America) for the polished ceramic sample.

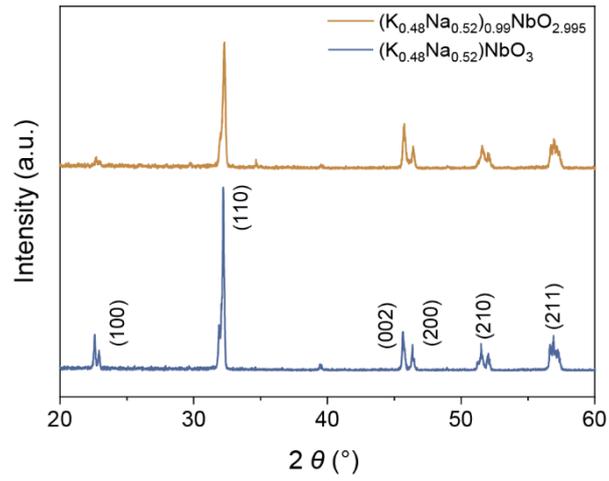

**Fig. S1.** XRD patterns of $(K_{0.48}Na_{0.52})_{0.99}NbO_{2.995}$ and $(K_{0.48}Na_{0.52})NbO_3$ ceramics.

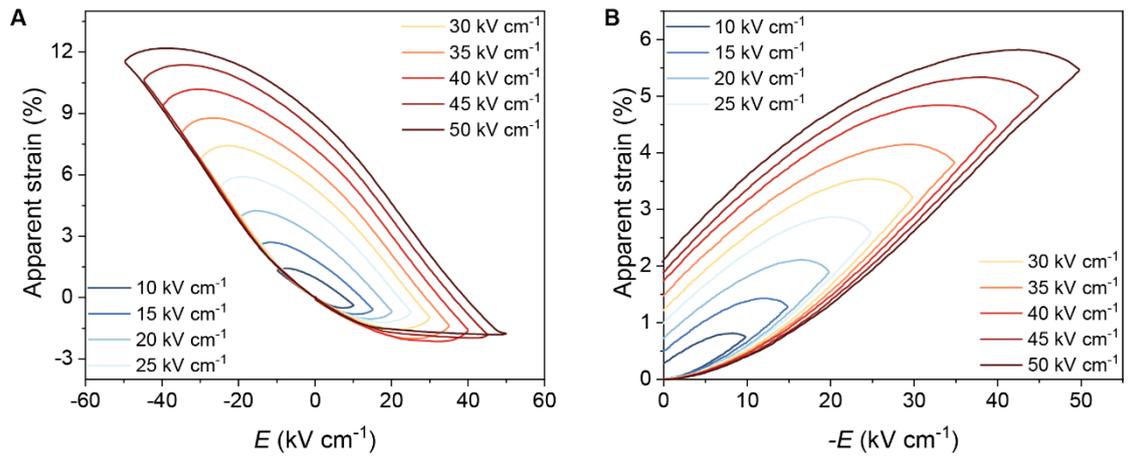

**Fig. S2.** (**A**) Bipolar and (**B**) unipolar *S-E* curves of KNN ceramics under different electric fields.

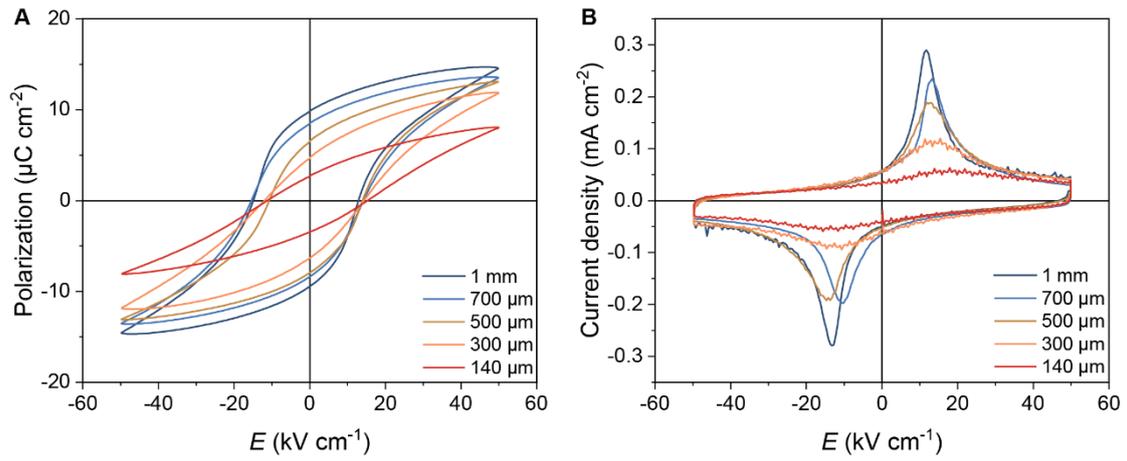

**Fig. S3.** (**A**) Polarization-electric field (*P-E*) loops and (**B**) current density-electric field (*J-E*) curves of KNN ceramics with different thicknesses.

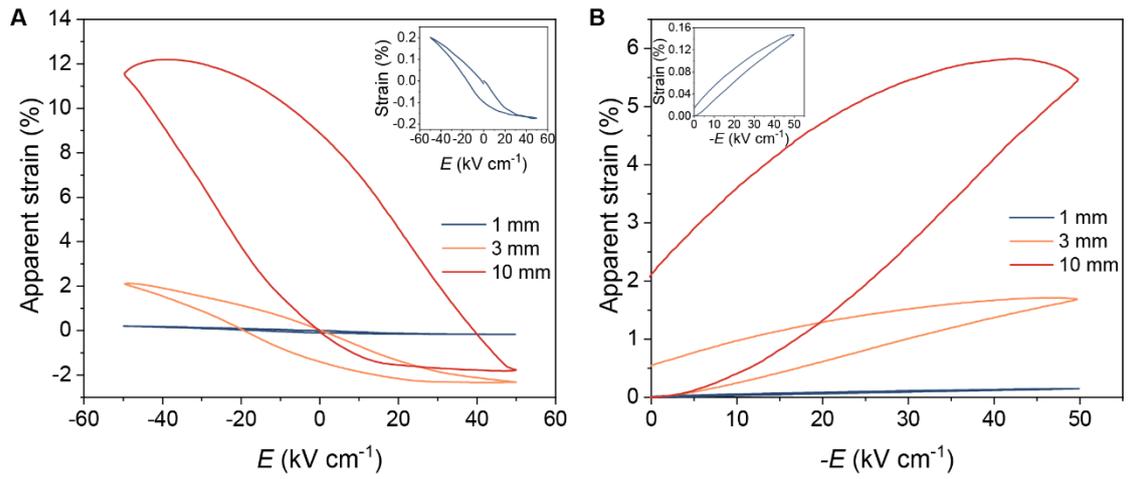

**Fig. *S4*.** (**A**) Bipolar and (**B**) unipolar *S-E* curves of KNN ceramics with a thickness of 140 μm measured by bottom sample holders with different diameters.

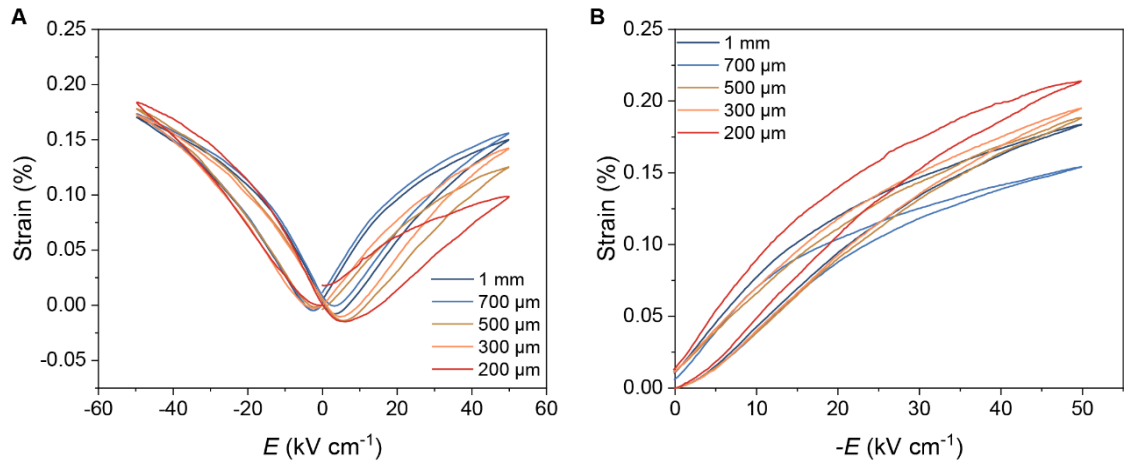

**Fig. S5.** (**A**) Bipolar and (**B**) unipolar *S-E* curves of BT ceramics with different thicknesses.

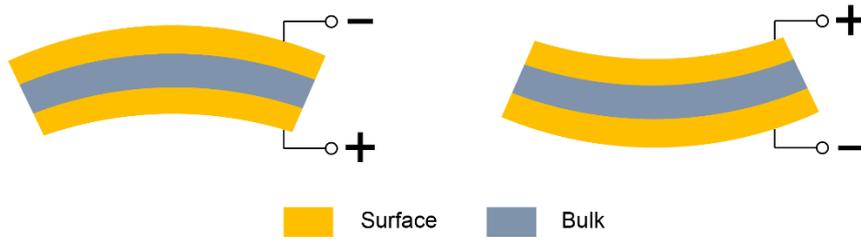

**Fig. S6.** Schematic diagram of bending direction of KNN ceramics under different electric field directions.

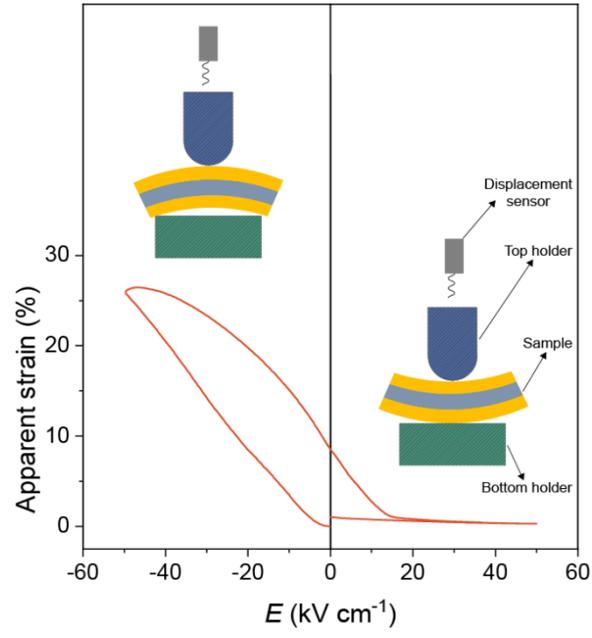

**Fig. S7.** Schematic of asymmetric bipolar *S-E* curve due to the effect of holder.

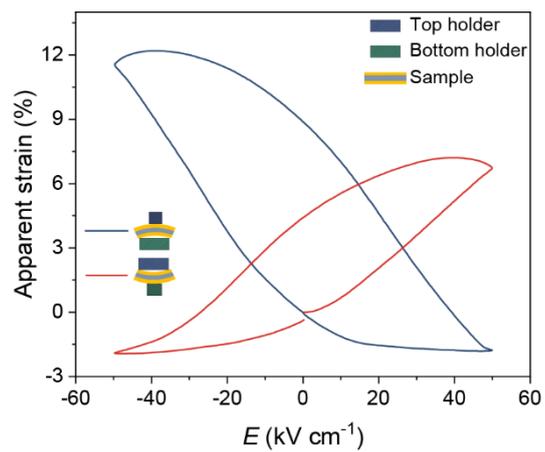

**Fig. S8.** *S-E* curves with different symmetrical directions measured by different holder configurations.

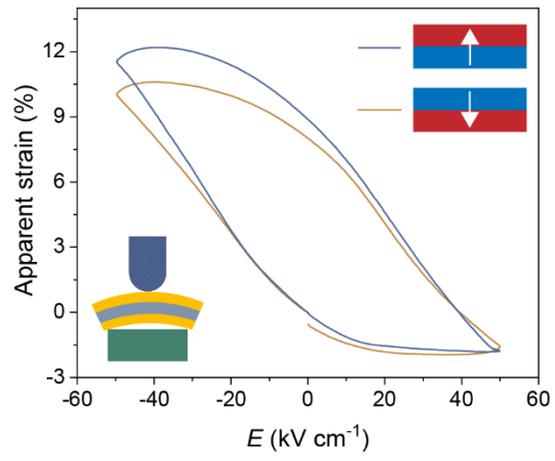

**Fig. S9.** *S-E* curves measured in different sample directions.

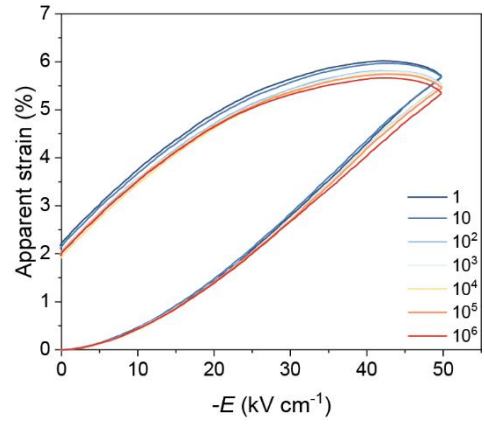

**Fig. S10.** Fatigue tests of unipolar *S-E* curves under an electric field of 50 kV cm$^{-1}$.

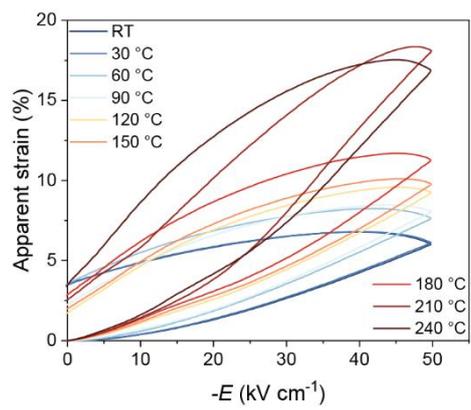

**Fig. S11.** Unipolar *S-E* curves with the temperature changing from 30 °C to 240 °C.

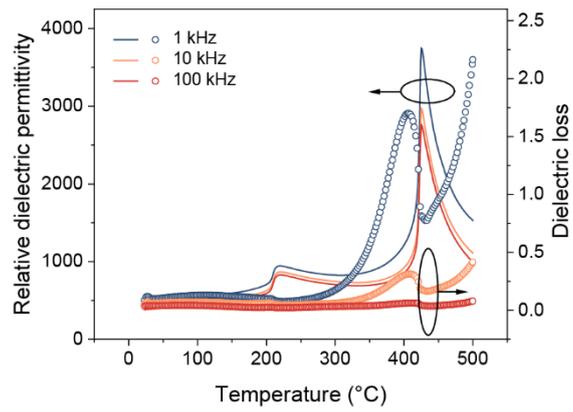

**Fig. S12.** Temperature dependent of relative dielectric permittivity and dielectric loss of KNN ceramics at 1 kHz, 10 kHz, and 100 kHz, respectively.